\documentstyle[12pt]{article}
\newcommand{\newc}{\newcommand}
\newc{\beq}    {\begin{equation}}
\newc{\eeq}    {\end{equation}}
\newc{\beqa}    {\begin{eqnarray}}
\newc{\eeqa}    {\end{eqnarray}}
\newc{\ba}    {\begin{array}}
\newc{\ea}    {\end{array}}
\newc{\st}    {\stackrel}
\newc{\f}    {\frac}

\def\NPB{{\em Nucl. Phys.} {\bf B}}

\def\PRL{\em Phys. Rev. Lett.}

\def\PRD{{\em Phys. Rev.} {\bf D} }

\begin{document}

\large
\normalsize
\noindent

\begin{titlepage}
\title{ RADIATION DAMPING  AT A BUBBLE WALL }
\author{ Jae-weon Lee, Kyungsub Kim,
and\\
Chul H. Lee,\\
      \it Department of Physics,
      \\ \it          Hanyang University,
                Seoul, 133-791, Korea
                \\and\\
Ji-ho Jang \\
\it Department of Physics,
      \\ \it          Seoul National  University,
                Seoul, 151-742, Korea
}

\setcounter{page}{1}
\date{}
\maketitle
\vspace{-9cm}
%\hfill KAIST-CHEP-100/95
\vspace{8cm}
PACS number(s): 12.15.Ji, 98.80.Cq
\vspace{1cm}

%\begin{abstract}

\hspace{.5cm}
The first order phase transition
proceeds via nucleation and growth of true
vacuum bubbles. When charged particles collide with
the bubble  they could radiate  electromagnetic
wave. We show that, due to an energy loss of the particles
by the radiation, the damping pressure
acting on the bubble wall  depends
on the velocity of the wall
even in a thermal equilibrium state.

%\end{abstract}
\vspace{4cm}
\maketitle

\end{titlepage}
\newpage

\vskip 5mm
There have been many studies on cosmological roles of
 first-order phase transitions which proceed via  nucleation and
growth of vacuum bubbles.
Studying  particle scatterings at a moving
bubble wall is important to know the bubble dynamics in a hot plasma.
For example, to calculate the velocity
of  electro-weak bubbles\cite{velocity,vel2} and
the CP violating charge transport rate  by the wall
available for baryogenesis,
\cite{kaplan}
one should know the reaction force acting on the
wall by the scattered particle({\it e.g.}  quarks, gauge bosons).

When charged particles collide with a bubble wall,
they could emit electromanetic radiation.
In this paper we study the effect of energy
being carried away by the radiation
on the pressure acting on the bubble wall.
When  particles enter a true vacuum bubble through a wall (say,
the electroweak bubble),  they could acquire a mass and are
decelerated. For example, a fermion $\psi$ may get a mass through the well known
Yukawa term $g\bar\psi\phi\psi=m \bar\psi \psi$, where $\phi$ is a Higgs field.
At this time if the particle is charged electomagnetically,
it can radiate  electromanetic wave due to the deceleration.
Let us calculate the pressure that
scattered particles exert on the bubble wall.
For simplicity we assume a linear profile for
the bubble wall.(See Fig.1.)
and choose a rest frame of the bubble wall.
Then,$m(x)=m_0 x/d$ ($0<x<d$) is the position dependent
 rest mass of the particle at the wall.
We also assume that the mean free path of the collision is much
shorter than the wall width $d$, hence WKB approximation
is good.
The radiation power of an accelerated particle is given by
the relativistic version
of Lamor's formula:
\beq
\frac{dE_{rad}}{dt}=\frac{\alpha}{m^2}\left (\frac{d\vec{k}}{dt} \right)^2,
\label{lamor}
\eeq
where $\alpha=2e^2/3c^3\simeq0.0611$ in natural units $\hbar=c=k=1$
and $\vec{k}=(k_x,k_y,k_z)$ is a 3-momentum of the particle.
Assuming that the wall is planar and parallel to the $y-z$ plane
we can treat the bubble as 1-dimensional one along the $x$-axis.
Energy, momentum and mass of the particle satisfy an usual relation
\beq
E^2\equiv m^2(x)+|\vec{k}|^2(x).
\eeq
Let us denote $k_x$ as $k$ from now on.
Differentiating  the above equation with time  $t$ and using
 $k=E dx/dt$, we get a force
\beq
\frac{dk}{dt}=-\frac{dm^2}{dx}\frac{1}{2E},
\label{dkdt}
\eeq
which is a starting point of the pressure calculation\cite{turok}.
However, if we also consider the energy carried  away by
 the radiation $E_{rad}$, then the total energy conserved
is $E_{tot}\equiv E_{rad}+E$ and the force  and
,hence, the pressure should be changed.
From $dE_{tot}/dt=0$ we obtain
\beq
\frac{k}{E}\frac{dk}{dt}+\frac{m}{E}\frac{dm}{dt}
+\frac{\alpha}{m^2} \left( \frac{dk}{dt} \right)^2=0,
\eeq
which has a solution
\beq
\frac{dk}{dt}=\frac{-m^3}{2\alpha E}\left [1-\sqrt{1-\frac{4\alpha E k}
{m^4}\frac{dm}{dt}}\right].
\eeq
Up to $O(\alpha)$ we can expand the root term and get
\beq
\frac{dk}{dt}\simeq -\frac{dm}{dx}\frac{m}{E}
- \frac{2\alpha}{kE}\left (\frac{dm}{dx} \right )^2.
\label{dk}
\eeq
Then the total pressure by
the collision of the particles in the plasma is
given by\cite{turok}
\beq
P=\int^{\infty}_{-\infty}{dx}
\int \frac{d^3k}{(2\pi)^3}\left [-\frac{dk}{dt}
f(E(k))\right ],
\eeq
where $f(E)=(Exp(\beta E)\pm 1)^{-1}$ is a distribution
function of  fermions and bosons respectively.
First, let us briefly review the well-known results
without the radiation damping( $\alpha=0$).
When the mean velocity of the plasma fluid $V$
(or minus of the bubble wall velocity relative to the fluid )
is zero, the first term of the Eq.(\ref{dk}) contributes
\beqa
P_1&=&\int^{\infty}_{-\infty}\frac{dm^2(x)}{dx} dx
\int \frac{d^3k}{(2\pi)^3}\frac{1}{2E}
\frac{1}{e^{\beta E}  \pm 1} \nonumber \\
&=& F(m,T)-F(0,T),
\eeqa
where $F(\phi,T)$ is  the $T$ dependent part of
free energy  at a temperature $T=\beta^{-1}$.
When $V\neq 0$ the distribution function
is changed to
\beq
f[\gamma(E-Vk)]=\left(e^{\beta \gamma(E-Vk)}\pm 1 \right)^{-1},
\eeq
where $\gamma$ is the Lorentz gamma factor.
However, using the fact that the phase factor $d^3k/E$ is Lorentz
invariant and changing the integration variable to $k'=\gamma (k-VE)$
and defining $E'\equiv \gamma (E-Vk)$
one can find that the $V$ dependency of $P_1$ disappears
\cite{turok}.
So one have needed to consider a non-equilibrium deviation of $f$
to calculate the velocity of the wall \cite{moore}.

Now let us consider the effect of the radiation,
 the second term of Eq.(\ref{dk}).
When $V=0$ the term contributes
\beq
P_2=2\alpha\int^{\infty}_{-\infty}\left( \frac{dm(x)}{dx} \right
)^2 dx
\int \frac{d^3k}{(2\pi)^3}\frac{1}{Ek(e^{\beta E}  \pm 1)}, \nonumber \\
\eeq
which vanishes, because the integrand is an odd function of k.
When $V\neq 0$
one can easily check that
due to the $1/k$ term the $V$ dependency survives
even under the change of the integration variable.
So in this case
\beqa
P_2 &=& 2\alpha\int^{\infty}_{-\infty}\left( \frac{dm(x)}{dx} \right
)^2 dx
\int \frac{d^3k}{(2\pi)^3}\frac{1}{Ek(e^{\beta \gamma(E-Vk)} \pm 1)}. \nonumber \\
&\equiv&2 \alpha\int^{\infty}_{-\infty}\left( \frac{dm(x)}{dx} \right
)^2 dx ~ I_2(x)
\label{p2}
\eeqa

% The authors are thankful to H. Kim
% for helpful discussions.
Let us discuss overall features of this integration qualitatively.
 To  investgate a singular behavior of the integrand
 in Eq.(\ref{p2}) near $k=0$
 we first consider  $k$ integration(along x-axis)
\beq
I\equiv \int^{\infty}_{-\infty} \frac{g(k)}{k} dk
=\int^{\infty}_{-\infty} \frac{g(k)-g(0)}{k-0} dk
+g(0)\int^{\infty}_{-\infty}\frac{1}{k} dk,
\eeq
where $g(k)=(E \{ Exp[\beta\gamma(E-Vk)]\pm 1\})^{-1}$.
The second term of the above equation
is zero.
For the first term,
since the contribution from  the integrand with $k \simeq 0$
dominates, one can approximate the term as
\beq
I\simeq \int^{k_0}_{-k_0} \frac{dg(k)}{dk} dk \simeq g(k_0)-g(-k_0),
\eeq
where $[k_0,-k_0]$ is a small momentum interval
in which the integrand contributes significantly.
If $V=0$  two terms  cancel each other.
When $V\ll 1$ one can expand them
up to $O(V)$ and expect that they  are  linearly dependent on $V$.
On the other hand, as $V\rightarrow 1 ~(\gamma \rightarrow
\infty)$ they are exponentially suppressed.
The integration over $k_y$ and $k_z$  are
just sums of $I$ with a shifted value of $E$, hence the overall behaviour
would not
be changed after 3-dimensional calculation.
( During the numerical study it is useful to
change the measure from $dk_ydk_z$ to $2\pi E dE$)
The numerical integration of Eq.(\ref{p2})
confirms this behaviour.(See Fig.2)

To be more concrete, let us obtain an approximate value
of the integration when $V\ll1$ for fermions.
In this case we can expand $f[\gamma(E-Vk)]\simeq f(E)-V\beta k f(E)[f(E)-1]
=f(E)+V\beta k f^{2}(E) Exp(-\beta E)$.
The integration with the first term gives zero
and the second term contributes
\beqa
I_2=V\beta \int \frac{d^3k}{(2\pi)^3}\frac{1}{E}f^{2}(E)
Exp(-\beta E)=\frac{(ln 2) T V}{2\pi^2}
\eeqa
, because
\beq
\int \frac{d^3k}{(2\pi)^3}\frac{1}{E}f^{2}(E)
Exp(-\beta E)\simeq \frac{(ln2) T^2}{2\pi^2},
\eeq
 to lowest oder in $(m/T)^2$ (See Ref.\cite{moore}).
Therefore, for the wall described in Fig.1, the pressure
by the radiation is
\beq
P_2=\frac{(ln 2)\alpha m_0^2 T }{\pi^2 d} V,
\eeq
which well fits with the numerical result
shown in Fig.2 when $V$ is small.
This pressure is proportional to the wall velocity
 up to the moderately relativistic case
  and exists even when  the system is in a thermal equilibrium.
Since $P_2$ is proportional to  the square of the charge,
an antiparticle contributes equally as a particle.

In summary we consider the radiation damping
of the particles colliding with  bubbles
and calculate the velocity dependent
pressure on the bubble wall.

\vskip 5.4mm
%\newpage
Authors are thankful to Dr. Myongtak Choi for
useful discussions.
This work was supported in part by the KOSEF
and in part by the  Korea research
foundation(BSRI-98-2441).

\newpage

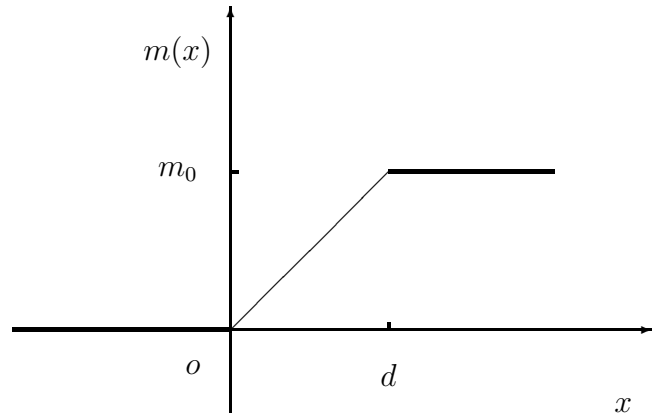
{\bf FIG.1.}\\
The mass of the particle $m(x)$ in the wall rest frame.

\vskip 3cm

{\bf FIG.2.}\\
The pressure by the radiation
damping of fermions colliding with the linear bubble wall
(Eq.(\ref{p2})).
Here $1/d=1=m_0=T$ for simplicity.

\newpage

\unitlength=1.00mm
\linethickness{0.8pt}
\begin{picture}(113.00,123.00)
\put(28.00,80.00){\vector(1,0){85.00}}
\put(57.00,69.00){\vector(0,1){54.00}}
%\put(30.00,106.00){\vector(1,0){10.00}}
%\put(40.00,102.00){\vector(-1,0){10.00}}
%\put(73.00,106.00){\vector(1,0){10.00}}
\linethickness{1.5pt}
\put(108.00,70.00){\makebox(0,0)[lc]{$x$}}
\put(50.00,117.00){\makebox(0,0)[cc]{$m(x)$}}
\put(50.00,101.00){\makebox(0,0)[cc]{$m_0$}}
%\put(50.00,100.00){\makebox(0,0)[cc]{$\triangle\theta$}}

%\put(35.00,110.00){\makebox(0,0)[cc]{$\Psi_I$}}
%\put(78.00,110.00){\makebox(0,0)[cc]{$\Psi_{II}$}}

\put(28.00,80.00){\line(1,0){29.00}}
\put(52.00,75.00){\makebox(0,0)[cc]{$o$}}
%\put(57.00,101.00){\line(1,0){42.00}}
\put(78.00,101.00){\line(1,0){22.00}}

\put(78.00,74.00){\makebox(0,0)[cc]{$d$}}
\put(60.00,45.00){\makebox(0,0)[cc]{$Fig.1.$}}
%\put(99.00,101.00){\line(0,0){0.00}}
%\put(99.00,101.00){\line(0,0){0.00}}
%\put(57.00,101.00){\line(0,-1){21.00}}
\put(57.00,80.00){\line(1,1){21.00}}
\linethickness{1.0pt}
\put(78.00,80.00){\line(0,1){1.00}}
\put(57,101.00){\line(1,0){1.00}}
\end{picture}

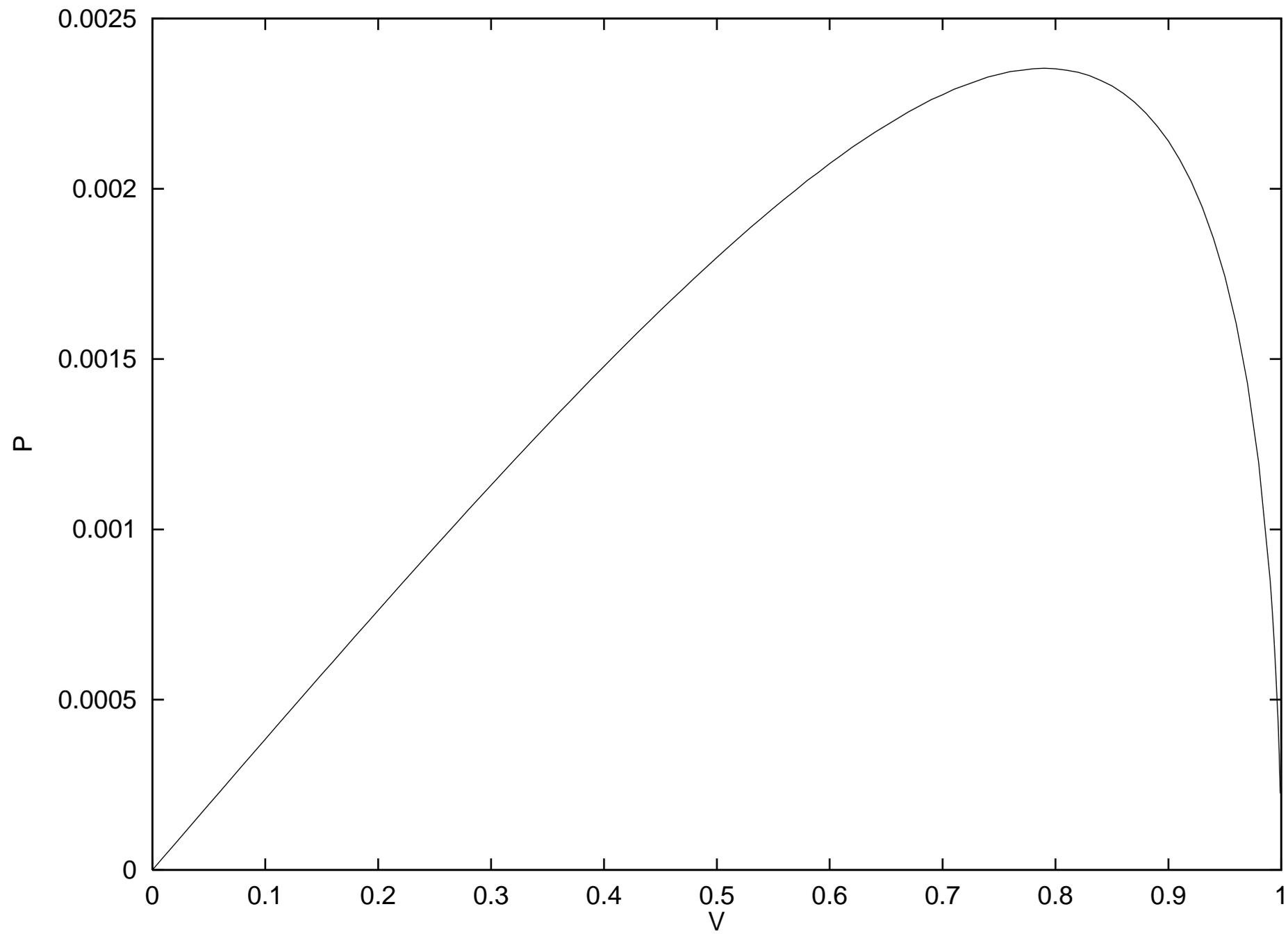
\end{document}